\documentclass[aps,prl,showpacs,twocolumn,amsmath,amssymb,superscriptaddress]{revtex4}
\usepackage{epsfig}
\usepackage{bm}
\def\be{\begin{equation}}
\def\ee{\end{equation}}
\def\la{\langle}
\def\ra{\rangle}
\begin{document}
\title{Leggett-Garg inequality with a kicked quantum pump}
\author{Andrew N. Jordan}
\affiliation{Department of Physics and Astronomy, University of Rochester, Rochester, New York 14627, USA}
\affiliation{D\'epartement de Physique Th\'eorique, Universit\'e de Gen\`eve,
        CH-1211 Gen\`eve 4, Switzerland}
\author{Alexander N. Korotkov}
\affiliation{
Department of Electrical Engineering, University of California,
Riverside, CA 92521-0204, USA}
\author{Markus B\"uttiker}
\affiliation{D\'epartement de Physique Th\'eorique, Universit\'e de Gen\`eve,
        CH-1211 Gen\`eve 4, Switzerland}
\date{October 28, 2005}
\begin{abstract}
A kicked quantum nondemolition measurement is introduced, where a
qubit is weakly measured by pumping current.  Measurement statistics
are derived for weak measurements combined with single qubit unitary
operations. These results are applied to violate a generalization of
Leggett and Garg's inequality. The violation is
related to the failure of the noninvasive detector assumption, and may
be interpreted as either intrinsic detector backaction, or the qubit
entangling the microscopic detector excitations.  The results are
discussed in terms of a quantum point contact kicked by a pulse
generator, measuring a double quantum dot.
\end{abstract}
 \maketitle An important goal in
the research of quantum phenomena in the solid state is to provide
realistic tests that demonstrate quantum behavior which no analogous
classical system could exhibit.  The best known example of such a
test is Bell's inequality (BI) \cite{bell}, but in submicron sized
samples the BI serves primarily as a test of entanglement rather
than ruling out local hidden variable theories \cite{peter}. The
seminal work of Leggett and Garg \cite{BIleggett} provides another
inequality involving only
one quantum variable together with a set of projective measurements.
This test demonstrates that the predictions of quantum mechanics are
incompatible with the philosophical assumptions of macrorealism and
a noninvasive detector.
An interesting parallel between the two inequalities is that the
role of hidden variables in the BI is played by trajectories in the
Leggett-Garg inequality (LGI). The belief that the quantum system
really takes a definite classical trajectory between two points
(chosen from an arbitrary probability distribution) may be disproved
with the LGI.

This Letter proposes a generalization of the LGI using
quantum nondemolition (QND) measurements weakly measuring the
quantum state by pumping current. Weak measurements, in contrast to
projective measurements, obtain partial information about the state
from an inherently noisy output, so wavefunction collapse happens
continuously. In the solid state, the typically weak coupling
between system and detector implies that weak measurements are the
norm. A generic problem arising in making a projective
measurement out of many weak measurements is that the quantum system
has its own Hamiltonian dynamics that effectively rotates the
measurement basis, preventing projective measurement.  The way around
this problem is with QND measurements.

\begin{figure}[b]
\begin{center}
\leavevmode
\psfig{file=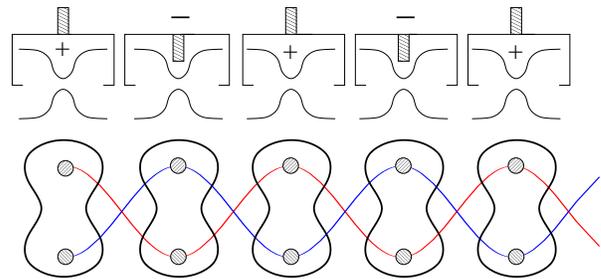,width=8cm}
\caption{(color online). Visualization of the kicked QND measurement.
A voltage pulse is applied to the QPC, followed by a
quiet period of zero voltage bias, lasting for a Rabi oscillation
period, followed by another pulse, and so on.  The up/down variation
is depicted, where the kicks come every half period, and the sign of
the voltage pulse alternates with every kick.  Read-out of the
coherent superposition of trajectories (red or blue) occurs by
measuring the sign of the current, and corresponds to an elementary
quantum pump.}
\label{fig2}
\end{center}
\vspace{-5mm}
\end{figure}
{\it Kicked QND.}--- The scheme we employ is that of kicked QND
measurements, introduced by Braginsky {\it et al.} and Thorne {\it
et al.} \cite{strobe} for the harmonic oscillator. In
Ref.~\cite{ANJ}, the idea is introduced for the two-state system by
two of the authors by making an analogy to a cat playing with a
string that moves in a circle: Rather than chasing the string
\cite{averin}, the cat sits in one spot waiting for the string to
come to it, and only then bats at it \cite{resonator}. The motion in
a circle comes from the evolution of a two-state
system, where $H= \Delta \sigma_x/2$ is the qubit Hamiltonian, and
$\Delta$ is the tunnel coupling of the symmetric qubit which defines
the Rabi oscillation period, $\tau_q=2\pi/\Delta$.  Although kicked
measurement may be realized in a wide variety of systems, we will
focus on a quantum point contact (QPC) kicked by a voltage pulse
generator.  This detector measures $\sigma_z$, the position of the
electron in a double quantum dot (DD) charge qubit as depicted in
Fig.~1.  The QPC detector is growing in experimental importance
\cite{DDexp1,DDexp2,DDexp3,DDexp4}, and Hayashi {\it et al.}
\cite{DDexp1} applied rectangular voltage pulses similar to the ones
we consider. 

An experimentally appealing variation on the idea of kicked
measurement is illustrated in Fig.~1, where a sequence of voltage
kicks of duration $\tau_V \ll \tau_q$ is applied to the QPC,
alternating in sign every {\it half} oscillation period.  The
parameters of the measurement process with an ideal QPC detector
\cite{bayesian,pilgram,gurvitz,AS} are specified by the currents,
$I_{1,2}$, that correspond to the different positions of the
electron in the DD, and the detector shot noise power $S_I =
eI(1-T)$ (where $T$ is the transparency).  The typical time needed
to distinguish the qubit signal from the background noise is the
measurement time $T_M = 4 S_I/(I_1-I_2)^2$.  If the qubit starts in
state $\vert 1 \ra$ [or $\vert 2 \ra$], so the kicks are in [or out
of] phase with the coherent oscillations, then the physical current
produced by the QPC is $(I_1-I_2)(\tau_V/\tau_q) >0$ [or
$(I_2-I_1)(\tau_V/\tau_q)<0$].  Thus, by simply determining the {\it
sign} of the current, a measurement of the quantum state can be
made.  Besides being a phase detector, this apparatus is also an
elementary quantum pump \cite{pump}, where the kicks provide one
time-changing parameter of zero average, and the intrinsic quantum
dynamics of the qubit provide the other changing parameter of zero
average that nevertheless causes a net flow of current \cite{note2}.

To characterize the result of each measurement kick, dimensionless
variables are introduced by defining the current origin at $I_0 =
(I_1+I_2)/2$ and scaling the current per pulse as $I-I_0 = x
(I_1-I_2)/2$, so $I_{1,2}$ are mapped onto $x=\pm 1$ (positive or
negative current in the pumping proposal).  The weak (static)
coupling between QPC and DD implies that $\tau_V \ll T_M$. In these
units, we take $x$ to be normally distributed with variance $D =
T_M/\tau_V \gg 1$.  The typical number of kicks needed to
distinguish the two states is $D$ kicks.  The measurement result
$\cal I$ after $N$ kicks is ${\cal I} = (1/N)\sum_{n=1}^{N} x_n$,
and we seek the conditional probability distribution $P({\cal I}, N
\vert \rho)$ of measuring the result ${\cal I}$, starting with a
given density operator $\rho$ prepared before the first kick. 
  The probability of
measuring the result $x_n$ after one kick is determined by the
state of the qubit just before the measurement, and is given by 
\be
P(x_n) = \rho_{11}^{(n)} P_1(x_n) + \rho_{22}^{(n)} P_2(x_n), 
\label{currentprob} \ee 
where $\rho_{ij}$ are the elements of the
density matrix in the $z$ basis, and the notation $P_{j}(x_n)$ is introduced for the
$j=1,2$ distributions of the $n$th kick.  These two distributions describe the detector 
output for the $n$th kick, if the electron resides only 
on one of the two dots.  The density matrix of the qubit is
updated based on information obtained from the measurement that just
occurred. This is done with the quantum Bayes rule \cite{bayesian}
that defines a non-unitary quantum map \cite{note3}:
\begin{eqnarray}
\hspace{-.4cm}\rho_{11}^{(n+1)}&=&1-\rho_{22}^{(n+1)}=
\frac{\rho_{11}^{(n)} P_1(x_n)}{\rho_{11}^{(n)}
P_1(x_n) + \rho_{22}^{(n)} P_2(x_n)},\nonumber  \\
\hspace{-.4cm}\rho_{12}^{(n+1)}&=&\left[\rho_{21}^{(n+1)}\right]^\ast
=\rho_{12}^{(n)}\sqrt{\rho_{11}^{(n+1)}
\rho_{22}^{(n+1)}/\rho_{11}^{(n)} \rho_{22}^{(n)}}.
\label{bayesrules}
\end{eqnarray}
The quantum Bayesian formalism provides additional insight into the
quantum detection process, and is well suited to analyze kicked QND
measurements. Equivalence with the quantum trajectories approach is
shown in Ref.~\cite{gmws} (see also Ref.~\cite{caves}). The
advantage of QND measurement in the quantum Bayesian approach
follows from using Eqs.~(\ref{bayesrules}) to calculate the
probability distribution $P({\cal I}, N \vert \rho)$ of current
${\cal I}$ after $N$ kicks, starting with the density matrix $\rho$:
\be P({\cal I}, N \vert \rho) = \rho_{11} P({\cal I},N \vert 1) +
\rho_{22} P({\cal I},N \vert 2), \label{qndanswer} \ee 
where $\rho_{11}, \rho_{22}$ are the diagonal matrix elements of the
original density matrix, and the
functions $P({\cal I},N|j),\ j=1,2$ are defined as Gaussian
probability distributions of the current, with average $(-1)^{j-1}$,
and variance $D/N$.  These two distributions describe the total detector output for $N$ kicks, 
if the electron resides only on one of the two dots.
In Eq.~(\ref{qndanswer}), the $N$ weak measurements simply compose to
give one $N$-times stronger measurement. As $N$ is increased, the
distributions limit to delta-functions giving either ${\cal I}=1,-1$
with probability $\rho_{11},\rho_{22}$ respectively.  A one-sigma
confidence is obtained when $N=D$ (see above).
The QND measurement output only involves the
diagonal density matrix elements, so the current output behaves {\it
exactly} as if it were simply collecting information about a
classical bit from a noisy process.
\begin{figure}[t]
\begin{center}
\leavevmode
\psfig{file=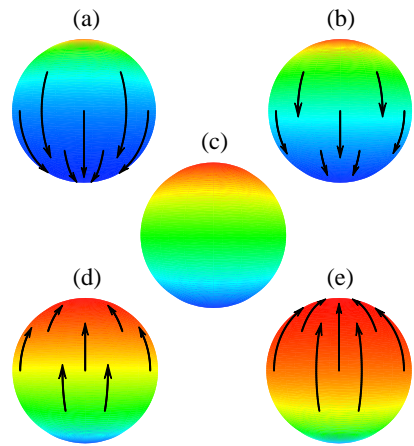,width=6cm}
\caption{(color online). The conditional evolution of all pure states
under kicked QND measurement is represented on the Bloch sphere. From
(a-e), the rescaled result of the measurement is $\gamma=(-1, -.5, 0,
.5, 1)$ respectively.  As the detector obtains more information, we
can with greater statistical certainty distinguish the
post-measurement quantum state, so the Bloch sphere is more and more
red ($\vert 1\ra$) or blue ($\vert 2 \ra$), depending on the value of $\gamma$ measured 
[color is assigned according to
$\rho_{11}-\rho_{22}$ of the final state].  The conditional evolution
of several representative states is also indicated with black arrows.
The North, South pole represent the states $\vert 1\ra$, $\vert 2\ra$.}
\label{fig3}
\end{center}
\vspace{-5mm}
\end{figure}
In spite of this fact, the Eqs.~(\ref{bayesrules}) allow us to deduce
the DD electron's density matrix prepared after the $N$ measurements
from our knowledge of ${\cal I}$: \be \rho' = \frac{1}{ \rho_{11}\,
e^\gamma+ \rho_{22}\, e^{-\gamma}}
\begin{pmatrix}
\rho_{11}\, e^\gamma & \rho_{12} \\
\rho_{12}^\ast   & \rho_{22}\, e^{-\gamma}
\end{pmatrix},
\label{rhonew}
\ee where $\gamma = {\cal I} N/D$ is the rescaled measurement result.
The conditional quantum dynamics of Eq.~(\ref{rhonew}) is illustrated
in Fig.~2 for all pure states, where $(x,y,z)$ are coordinates on the
Bloch sphere.  The $x$ and $y$ behavior follows from $z$, which is in
turn conditioned on the detector output ${\cal I}$, so the sphere is
colored according to the conditional evolution of $z$.  If $\gamma$ is
positive (negative), then states are ``attracted'' toward the North
(South) pole.  As $\gamma$ grows increasingly positive or negative, we
become more confident which state the qubit has collapsed to, so the
sphere is more and more red ($\vert 1\ra$) or blue ($\vert 2 \ra$),
but notice that this depends on the initial state.  The conditional
evolution of several representative states is indicated with black
arrows.

{\it Generalized LGI.}--- While the point of the kicked QND
proposal was to effectively turn off the qubit unitary evolution while
the measurement is taking place, kicked measurement provides a simple
way of generating a single-qubit rotation: Waiting some fraction $r$
of a Rabi oscillation between kicks defines a phase shift $\phi =
2 \pi r$ on the DD qubit.  Consider now an experiment, comprised of $N_1$
kicks, followed by a single qubit unitary operation ${\bf U}(\phi)$,
followed by $N_2$ kicks.  The measurement results are defined as
${\cal I}_1=(1/N_1)\sum_{n=1}^{N_1} x_n, \; {\cal
I}_2=(1/N_2)\sum_{n=N_1+1}^{N_1+N_2} x_n$.  We seek the normalized
probability distribution $P({\cal I}_1; {\cal I}_2)$ of finding
current ${\cal I}_1$ after $N_1$ kicks, and ${\cal I}_2$ after $N_2$
subsequent kicks.  This distribution may also be interpreted as a
``joint counting statistics''.  After the first $N_1$ kicks, the
measured current ${\cal I}_1$ will occur with a probability density
given by (\ref{qndanswer}), and prepares a post-measurement density
matrix $\rho'$ (\ref{rhonew}).  The subsequent unitary operation
rotates this density matrix, $\rho^{\rm new} = {\bf
U}\, \rho'\, {\bf U}^\dagger$.  The following set of $N_2$ kicks start
with $\rho^{\rm new}$ and continue to measure in
the $z$-basis as before.  Equation (\ref{qndanswer}) may be applied
again to obtain
\begin{eqnarray}
P({\cal I}_1; {\cal I}_2) &=& [\rho_{11} P({\cal I}_1,N_1 \vert 1) + \rho_{22} P({\cal I}_1,N_1
  \vert 2)]  \label{oneshift} \\
  &\times&  [\rho^{\rm new}_{11} P({\cal I}_2,N_2 \vert 1) +
\rho^{\rm new}_{22} P({\cal I}_2,N_2 \vert 2)],      \nonumber
\end{eqnarray}
where the new density matrix elements are
\begin{eqnarray}
\rho^{\rm new}_{11} &=& [\cos^2(\phi/2)
\rho_{11} e^\gamma + \sin^2 (\phi/2)
\rho_{22} e^{-\gamma} \nonumber \\ &-&  \sin \phi \; {\rm
  Im}\rho_{12} ]/( \rho_{11} e^\gamma + \rho_{22} e^{-\gamma}),  \nonumber \\
\rho^{\rm new}_{12} &=&  [ {\rm Re} \rho_{12}+ (i/2) \sin\phi\;
(\rho_{11} e^\gamma -\rho_{22} e^{-\gamma}) \nonumber \\
&+& i \cos\phi \; {\rm Im} \rho_{12}]/(\rho_{11} e^\gamma + \rho_{22}
e^{-\gamma}).
\label{rhonew2}
\end{eqnarray}
Note that the outcome of the first $N_1$ kicks, ${\cal I}_1$, appears
in the expression involving the second set of kicks, so the
distribution does not factorize.  It is straightforward to generalize
the results (\ref{oneshift},\ref{rhonew2}) to any number of
dislocations in the pulse sequence, each of which has a phase shift.

We now demonstrate how these results may be used to violate a
generalized LGI.  A generalized LGI has been discussed by Ruskov
{\it et al.} \cite{BItime} for the current correlations and the
spectral noise peak generated by a qubit. Our setup has the
advantage of full tunability of phase-shifts and measurement
strength and thus permits a LGI test over a wide range of
parameters.  The original proposal \cite{BIleggett} derived an
inequality involving correlation functions from three experiments,
each consisting of two projective measurements done at specified
times starting from the same initial condition.  The beauty of weak
measurements is that the inequality may be violated with only {\it
one} set of measurements together with statistical averaging.  To
derive the weak measurement generalization of the LGI, consider
three kicks surrounding two phase shifts $\phi_{1,2}$. Define the
correlation function $B = S_{12} + S_{23} - S_{13}$, where
$S_{nm}=\la {\cal I}_n {\cal I}_m \ra$; $n,m=1,2,3$.  The assumptions of
``macrorealism and a noninvasive detector'' \cite{BIleggett} are
introduced with a white, additive, noise model of the
detection process (characteristic of QPC electron transport).
The measured result ${\cal I}_n$ can be decomposed into a system signal 
and detector noise contribution, ${\cal I}_n = C_n + \xi_n$.  
The signal $C_n$ describes the DD state at measurement $n$, 
while the detector noise term, $\xi_n$, is white
Gaussian noise (discussed previously), of zero average and variance $\la \xi_n^2\ra = D/N_n$. 
The signal contribution $C_n$ may be endowed with classical hidden variables
$\{ \lambda \}$, chosen from any probability distribution.
The signal can now change arbitrarily between
measurements, but only in a bounded way, $-1 \le C_n(\{\lambda\}) \le 1$.
The noninvasive detector assumption implies that the detector noise
does not affect the measured system in the past or the future, so
$\la \xi_n C_m(\{\lambda\})\ra =0$, for any $n,m$.
These assumptions imply that $S_{nm} = \la C_n(\{\lambda\}) C_m(\{\lambda\}) \ra_r$, 
where $\la \ldots \ra_r$ denotes further averaging over the hidden variables
$\{\lambda\}$, as well as over realizations or initial conditions.  From 
the bound on each of the signal contributions, it is straightforward to 
show that $B \le 1$, concluding the weak
measurement generalization of the LGI.

Starting with any DD electron state, we find quantum mechanically
from the generalization of (\ref{oneshift},\ref{rhonew2}) that
\begin{eqnarray} B &=& \cos \phi_1 +\cos \phi_2 - \cos \phi_1\cos
\phi_2 \label{bellineq} \\
&+& \sin \phi_1\sin \phi_2 \exp(- N_2 /2D),    \nonumber
\end{eqnarray}
for an arbitrary number of kicks $N_1, N_2, N_3$ made around the
phase shifts.  The first three terms in (\ref{bellineq}) cannot
violate the LGI, and it is the last term that is responsible for the
violation. In the weak measurement limit, $N_2 \ll D$, the Bell-like
parameter takes the form, $B \approx \cos \phi_1 +\cos \phi_2 - \cos
(\phi_1 +\phi_2)$, and is maximally violated for $\phi_1 =\phi_2 =
\pi/3$ so $B = 3/2$.  The physical interpretation for the
suppression of the critical term in (\ref{bellineq}) is the
following: If measurement $2$ had not been made, the system travels
in a coherent superposition of trajectories (red and blue in Fig.~1)
between $1 \rightarrow 3$.  The intermediate measurement gives the
necessary third point, but also yields information (at a rate
$D^{-1}$ per kick) about which trajectory the quantum system
``really'' took \cite{whichpath}.  This information manifests itself
in making it harder to violate the LGI. In the projective
measurement limit, $N_2 \gg D$, we are statistically confident which
trajectory the system took, and it becomes impossible to violate the
LGI.

An alternative picture may be seen by reconsidering three incident
electron groups on the QPC, spaced by a phase shift $\phi_{1,2}$ on
the DD qubit.  Rather than directly project the QPC electrons after
each passes (as is necessary for the quantum Bayesian approach), we
use a well known property of quantum circuits, that the predictions
of quantum mechanics are identical if the projective measurements
are delayed to the end of the all unitary operations.  Then the
above procedure is identical to the quantum circuit drawn in Fig.~3,
where each initial left scattering state $\vert L\ra$ encodes many
transport electrons.  Rather than attribute the correlations
(\ref{bellineq}) to detector backaction, another interpretation is
to see the above procedure as the DD qubit effectively  creating
entanglement between the transport electrons.  For simplicity, we
consider the $1/2$ transparency point (see Ref.~\cite{pilgram}
for a more general discussion).  Following the detector treatment in
Refs. \cite{gurvitz,AS}, the transmission (reflection)
amplitudes $t_{1,2} (r_{2,1}) = \sqrt{1/2 \pm \epsilon}$ of the two
scattering matrices corresponding to the two positions of the DD
electron are expanded in the detector sensitivity $\epsilon$ to
second order. If the qubit is in state $\vert 1 \ra, \vert 2\ra$
then the out-going detector scattering states are $\vert \chi_{1,2}
\ra = \vert s \ra (1-\epsilon^2/2) \pm \epsilon \vert a \ra$, where
$\vert s\ra, \vert a \ra = (\vert L\ra \pm \vert R \ra)/\sqrt{2}$
are combinations of the left/right scattering states.  Before
measurements are made, the state is given by
\begin{eqnarray}
\vert \Psi \ra &=& {\bf U}(\phi_1){\bf U}(\phi_2) \vert \psi \ra \vert 0\ra
(1-3\epsilon^2/2) \nonumber  + \epsilon \vert \psi'\ra
[\cos \phi_+ \nonumber \\ &\times&
 (\vert 1\ra + \vert 3\ra) \nonumber + \cos\phi_- \vert 2\ra]
+\epsilon \vert \psi''\ra [\sin
\phi_+ (\vert 3\ra -\vert 1\ra)
\nonumber \\ &+&
 \sin\phi_- \vert 2\ra] + \epsilon^2
\vert \psi \ra [\cos \phi_- (\vert 1'\ra +\vert 3'\ra) + \cos\phi_+
\vert 2'\ra]
\nonumber \\ &+&
\epsilon^2 \vert \psi'''\ra [\sin \phi_- (\vert 3'\ra
-\vert 1'\ra) +  \sin\phi_+ \vert 2'\ra],
 \label{state}
\end{eqnarray}
where $\vert 0 \ra = \vert sss \ra$, $\vert 1,2,3 \ra= \vert ass, sas,
ssa \ra$, $\vert 1',2',3'\ra =\vert saa,asa,aas \ra$, and $\vert klm
\ra \equiv \vert k\ra_I\vert l\ra_{II}\vert m\ra_{III}$.  The DD
states are defined as $\vert \psi, ','','''\ra = \alpha \vert
1 \ra +\beta \vert 2\ra, -\alpha \vert 1 \ra +\beta \vert 2\ra, i
(\beta \vert 1 \ra - \alpha \vert 2\ra), i (\beta \vert 1 \ra + \alpha
\vert 2\ra) $, and $\phi_{\pm} = (\phi_1 \pm \phi_2)/2$.  The first
dominant term is separable and alone can produce no correlations,
while the remaining terms are entangled.  Using projection operators
on the right scattering states in order to calculate current
correlators recovers the weak measurement limit of (\ref{bellineq}).
\begin{figure}[t]
\begin{center}
\leavevmode
\psfig{file=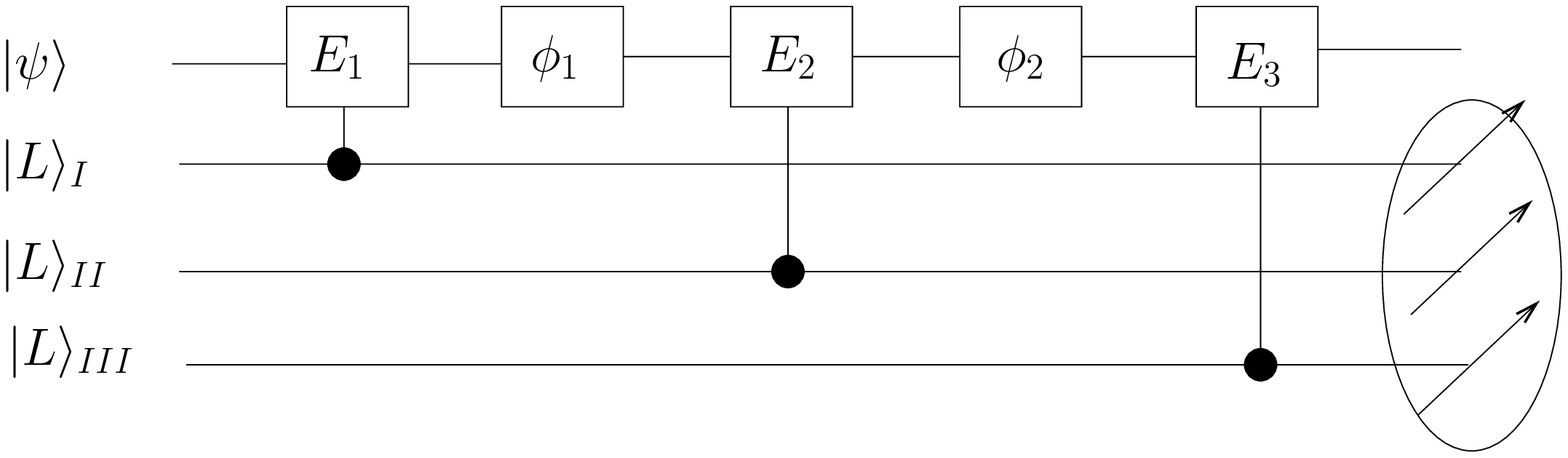,width=8cm}
\caption{Quantum circuit analysis of the generalized LGI:
Three sets of initially unentangled QPC electrons in the left
scattering states $\vert L\ra_{I,II,III}$ are entangled with the DD electron
state $\vert \psi \ra$ with operations $E_{1,2,3}$.  The DD electron undergoes
two different phase shifts $\phi_{1,2}$ between the passing electron
bunches.  Projective measurements are made on the QPC electrons in the
current collector, and are delayed until the end of all unitary
operations in the entanglement picture. }
\label{fig4}
\end{center}
\vspace{-5mm}
\end{figure}

{\it Conclusions.}--- We have proposed a kicked qubit readout scheme that
is both a quantum nondemolition measurement and a quantum pump.
Kicked measurements combined with unitary operations were used to
formulate and violate a weak measurement generalization of Leggett and
Garg's inequality.  
The fact
that our proposal uses one set of pulses to accomplish both weak
measurements and phase shifts provides an important advantage for an
experiment aimed at violating the LG inequality.

We thank L. P. Kouwenhoven, E. V. Sukhorukov, R. Ruskov, and B. Trauzettel
for helpful discussions. This work was supported by MaNEP and the SNF
(A.N.J. and M.B.) and the NSA/ARDA/ARO (A.N.K.).


\vspace{-5mm}
\begin{thebibliography}{05}
\vspace{-5mm}
\bibitem{bell}
J.~S. Bell, Physics {\bf 1}, 195 (1964).

\bibitem{peter}
P. Samuelsson, E.~V. Sukhorukov, and M. B\"uttiker,
Phys. Rev. Lett. {\bf 92}, 026805 (2004);
A.~V. Lebedev, G.~B. Lesovik, and G. Blatter, Phys. Rev. B {\bf 71},
045306 (2005); C.~W.~J. Beenakker, cond-mat/0508488 (2005).

\bibitem{BIleggett}
A.~J. Leggett and A. Garg,
Phys. Rev. Lett. {\bf 54}, 857 (1985);
A.~J. Leggett, J. Phys.: Condens. Matter {\bf 14}, R415 (2002).

\bibitem{strobe}
V.~B. Braginsky, Yu.~I. Vorontsov, and F.~Ya. Khalili,
JETP Lett. {\bf 27}, 276 (1978);
K.~S. Thorne {\it et al.},
Phys. Rev. Lett. {\bf 40}, 667 (1978).

\bibitem{ANJ}
A.~N. Jordan and M. B\"uttiker,
Phys. Rev. B {\bf 71}, 125333 (2005).

\bibitem{averin}
D.~V. Averin,
Phys. Rev. Lett. {\bf 88}, 207901 (2002).

\bibitem{resonator}
R. Ruskov, K. Schwab, and A.~N. Korotkov,
Phys. Rev. B {\bf 71}, 235407 (2005).

\bibitem{DDexp1}
T. Hayashi {\it et al.},
Phys. Rev. Lett. {\bf 91}, 226804 (2003).

\bibitem{DDexp2}
J.~M. Elzerman {\it et al.},
Phys. Rev. B {\bf 67}, R161308 (2003).

\bibitem{DDexp3}
J.~R. Petta  {\it et al.},
Phys. Rev. Lett. {\bf 93}, 186802 (2004).

\bibitem{DDexp4}
A.~K. H\"uttel {\it et al.},
Phys. Rev. B {\bf 72}, R081310 (2005).

\bibitem{gurvitz}
S.~A. Gurvitz,
Phys. Rev. B {\bf 56}, 15215 (1997).

\bibitem{bayesian}
A.~N. Korotkov, Phys. Rev. B {\bf 60}, 5737 (1999);
Phys. Rev. B {\bf 63} 115403 (2001).

\bibitem{pilgram}
S. Pilgram and M. B\"uttiker,
Phys. Rev. Lett. {\bf 89}, 200401 (2002);
A.~A. Clerk, S.~M. Girvin, and A.~D. Stone,
Phys. Rev. B {\bf 67}, 165324 (2003).

\bibitem{AS}
D.~V. Averin and E.~V. Sukhorukov,
Phys. Rev. Lett. {\bf 95}, 126803 (2005).

\bibitem{pump}
P.~W. Brouwer, Phys. Rev. B {\bf 58}, R10135 (1998);
J.~E. Avron {\it et al.},
Phys. Rev. B {\bf 62}, R10618 (2000);
M. Moskalets and M. B\"uttiker,
Phys. Rev. B {\bf 70}, 245305 (2004).

\bibitem{note2} Eventually random bit-flip errors occur
if the kicks are imperfect delta-functions, causing the pumped
current to vanish asymptotically \cite{ANJ}. A sustainable quantum pump may be
built with a simple feedback loop;
R. Ruskov and A.~N. Korotkov, Phys. Rev. B {\bf 66}, R041401 (2002).

\bibitem{note3} This is equivalent to weak entanglement of the QPC
electrons with the DD qubit, with projective measurement on the QPC
electrons, executing a POVM on the DD qubit; A.~N. Jordan and
A.~N. Korotkov, cond-mat/0606676.

\bibitem{gmws}
H.-S. Goan {\it et al.}, Phys. Rev. B {\bf 63}, 125326 (2001).

\bibitem{caves}
C. M. Caves {\it et al.}, Phys. Rev. A {\bf 65} 022305 (2002).


\bibitem{BItime}
R. Ruskov, A.~N. Korotkov, and A. Mizel, Phys. Rev. Lett. 
{\bf 96}, 200404 (2005).

\bibitem{whichpath}
E. Buks {\it et al.},
Nature {\bf 391}, 871 (1998).

\end{thebibliography}
\end{document}